\documentclass[aps,superscriptaddress]{revtex4-1}
\usepackage{graphicx}
\usepackage{setspace}  
\usepackage{amsmath}
\usepackage{amssymb} 

\newcommand{\rrra}{${| 
\put(02,0){{\makebox(0,0){$\Diamond$}}}
\put(07,0){{\makebox(0,0){$\Diamond$}}}                                                                                  
\put(012,0){{\makebox(0,0){$\Diamond$}}}                                                                                  
~~~~\rangle                                                                                                              
}$
}

\newcommand{\rrrb}{${| 
\put(02,0){{\makebox(0,0){$\Diamond$}}}
\put(07,0){{\makebox(0,0){$\Diamond$}}}                                                                                  
\put(012,0){{\makebox(0,0){$\Diamond$}}}                                                                                  
\put(04.5,05){{\makebox(0,0){$\Diamond$}}}                                                                                  
~~~~\rangle                                                                                                              
}$
}

\newcommand{\rrrc}{${| 
\put(02,0){{\makebox(0,0){$\Diamond$}}}
\put(07,0){{\makebox(0,0){$\Diamond$}}}                                                                                  
\put(012,0){{\makebox(0,0){$\Diamond$}}}                                                                                  
\put(09.5,05){{\makebox(0,0){$\Diamond$}}}                                                                                  
~~~~\rangle                                                                                                              
}$
}

\newcommand{\rrrd}{${| 
\put(02,0){{\makebox(0,0){$\Diamond$}}}
\put(07,0){{\makebox(0,0){$\Diamond$}}}                                                                                  
\put(012,0){{\makebox(0,0){$\Diamond$}}}                                                                                  
\put(09.5,05){{\makebox(0,0){$\Diamond$}}}                                                                                  
\put(04.5,05){{\makebox(0,0){$\Diamond$}}}                                                                                  
~~~~\rangle                                                                                                              
}$
}

\newcommand{\rrre}{${|
\put(02,0){{\makebox(0,0){$\Diamond$}}}
\put(07,0){{\makebox(0,0){$\Diamond$}}}                                                                                  
\put(012,0){{\makebox(0,0){$\Diamond$}}}                                                                                  
\put(09.5,05){{\makebox(0,0){$\Diamond$}}}                                                                                  
\put(04.5,05){{\makebox(0,0){$\Diamond$}}}                                                                                  
\put(07,10){{\makebox(0,0){$\Diamond$}}}                                                                                  
~~~~\rangle 
}$
}

\newcommand{\wwwa}{${| 
\put(07,0){{\makebox(0,0){$\Diamond$}}}                                                                                  
\put(012,0){{\makebox(0,0){$\Diamond$}}}                                                                                  
~~~~\rangle                                                                                                              
}$
}

\newcommand{\wwwb}{${| 
\put(07,0){{\makebox(0,0){$\Diamond$}}}                                                                                  
\put(012,0){{\makebox(0,0){$\Diamond$}}}                                                                                  
\put(04.5,05){{\makebox(0,0){$\Diamond$}}}                                                                                  
~~~~\rangle                                                                                                              
}$
}

\newcommand{\wwwc}{${| 
\put(07,0){{\makebox(0,0){$\Diamond$}}}                                                                                  
\put(012,0){{\makebox(0,0){$\Diamond$}}}                                                                                  
\put(09.5,05){{\makebox(0,0){$\Diamond$}}}                                                                                  
~~~~\rangle                                                                                                              
}$
}

\newcommand{\wwwd}{${| 
\put(07,0){{\makebox(0,0){$\Diamond$}}}                                                                                  
\put(012,0){{\makebox(0,0){$\Diamond$}}}                                                                                  
\put(09.5,05){{\makebox(0,0){$\Diamond$}}}                                                                                  
\put(04.5,05){{\makebox(0,0){$\Diamond$}}}                                                                                  
~~~~\rangle                                                                                                              
}$
}

\newcommand{\wwwe}{${|
\put(07,0){{\makebox(0,0){$\Diamond$}}}                                                                                  
\put(012,0){{\makebox(0,0){$\Diamond$}}}                                                                                  
\put(09.5,05){{\makebox(0,0){$\Diamond$}}}                                                                                  
\put(04.5,05){{\makebox(0,0){$\Diamond$}}}                                                                                  
\put(07,10){{\makebox(0,0){$\Diamond$}}}                                                                                  
~~~~\rangle 
}$
}

\newcommand{\wwwf}{${|
\put(07,0){{\makebox(0,0){$\Diamond$}}}                                                                                  
\put(012,0){{\makebox(0,0){$\Diamond$}}}                                                                                  
\put(09.5,05){{\makebox(0,0){$\Diamond$}}}                                                                                  
\put(04.5,05){{\makebox(0,0){$\Diamond$}}}                                                                                  
\put(07,10){{\makebox(0,0){$\Diamond$}}}
\put(04.5,14.5){{\makebox(0,0){$\Diamond$}}}                                                                                                                                                                    
~~~~\rangle 
}$
}

\newcommand{\be}{\begin{equation}}
\newcommand{\ee}{\end{equation}}

\newcount{\myi}
\newcommand{\Repeat}[2]{%
    \myi=0
    \loop
        \ifnum\myi<#2
        #1
        \advance\myi by 1
    \repeat
}

\begin{document}

\title{Avalanches in the Raise and Peel model in the presence of a wall}
\author{Edwin Antillon}
\affiliation{Departments of Physics,  Purdue University,  West Lafayette, IN 47907}
\author{Birgit Wehefritz-Kaufmann}
\affiliation{Departments of Physics,  Purdue University,  West Lafayette, IN 47907}
\affiliation{Departments of Mathematics,  Purdue University,  West Lafayette, IN 47907}
\author{Sabre Kais}
\affiliation{Departments of Physics,  Purdue University,  West Lafayette, IN 47907}
\affiliation{Department of Chemistry, Purdue University, West Lafayette, IN 47907}
\affiliation{Qatar Environment and Energy Research Institute, Doha, Qatar}

\begin{abstract}
 We investigate a non-equilibrium one-dimensional model known as the raise and peel model describing 
  a growing surface which grows locally and has non-local desorption. For specific values of adsorption ($u_a$) and 
  desorption ($u_d$) rates the model shows interesting features. At $u_a = u_d$, the model is described by a conformal field
  theory (with conformal charge $c=0$) and its stationary probability can
  be mapped to the ground state of the XXZ quantum chain. Moreover, for
  $u_a \geq  u_d$, the model shows a  phase in which the the
  avalanche distribution is scale invariant. 
  In this work we study the surface dynamics by looking at avalanche distributions using Finite-size Scaling formalism 
  and explore the effect of adding a wall to the model. The model shows the same universality 
  for the cases with and without a wall for an odd number of tiles removed,
   but we find a new exponent in the presence of a wall for an even number of avalanches released. 
   We provide new conjecture for the probability distribution of avalanches with a wall obtained by using exact 
   diagonalization of small lattices and Monte-Carlo simulations.
\end{abstract}

\maketitle

\section{Introduction}
The Raise and Peel Model (RPM) is a Markov process first proposed by \cite{gier} describing the evolution of a growing surface with a fluctuating interface in one dimension. This model has been found to belong to a new universality class
 in non-equilibrium phenomena \cite{gier,gier1,Alcaraz0,alcaraz,alcaraz_facets}.
For a particular value of the adsorption ($u_a$) and desorption ($u_d$)
rates, the model exhibits a phenomenon of 
self-organized criticality \cite{bak,dhar} where probability distributions of desorption events 
show long tails and are characterized by a varying critical exponent that depends on a single parameter 
given by the ratio of the adsorption and desorption rates \cite{alcaraz,alcaraz_facets}. 

When adsorption and desorption rates are equal, the model becomes solvable. 
This goes back to a connection established by Razumov and Stroganov \cite{RS,RStwisted,gier2} 
which relates the two-dimensional dense $O(n=1)$ fully packed loop models 
(enumerating the stationary state probability distributions of RPM)  
to those a ground state wavefunctions of the XXZ chain with L sites
\cite{alcaraz_facets,Batchelor,alcaraz_spin}. 
 Moreover, the spectra can be obtained by conformal field theory with charge c=0 \cite{Knops,Henkel}.
This offers a nice mathematical structure, which  allows to make conjectures using small lattices
for expressions of physical quantities that remain valid for quantities for any system size.

In this work, we study the effect on avalanches in the presence of a wall (RPMW) since little is known about
this effect when the boundary is allowed to fluctuate. Some other interesting results with the wall have been 
reported for example in \cite{Alcaraz2,Pyatov}. 
In Section~\ref{sec:RPM} we describe the stochastic rules for the model 
with and without a wall and highlight some of the known results for these two cases. 
In Section~\ref{sec:spectra}, we compare the energy spectra of the stochastic Hamiltonian of the XXZ quantum chain
for different spin sectors. Lastly in Section~\ref{sec:avalanches},
 we compute critical exponents for avalanche distributions for RPM and RPMW and derive 
new conjectures for probability expression with a wall.

\section{Raise and Peel Models}\label{sec:RPM}
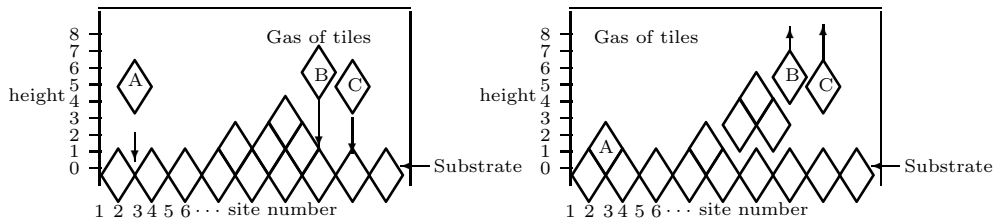
\begin{figure}[t]
  \begin{tabular}{cc}
\setlength{\unitlength}{0.180900pt}
\ifx\plotpoint\undefined\newsavebox{\plotpoint}\fi
\sbox{\plotpoint}{\rule[-0.200pt]{0.400pt}{0.400pt}}%
\begin{picture}(750,450)(0,0)
\sbox{\plotpoint}{\rule[-0.200pt]{0.400pt}{0.400pt}}%
\put(50.0,40.0){\rule[-0.200pt]{0.400pt}{66pt}}
\put(700.0,40.0){\rule[-0.200pt]{0.400pt}{66pt}}
\put(30,75){\rule[-0.200pt]{4.818pt}{0.400pt}}
\put(30.0,110){\rule[-0.200pt]{4.818pt}{0.400pt}}
\put(30.0,145){\rule[-0.200pt]{4.818pt}{0.400pt}}
\put(30.0,180){\rule[-0.200pt]{4.818pt}{0.400pt}}
\put(30.0,215){\rule[-0.200pt]{4.818pt}{0.400pt}}
\put(30.0,250){\rule[-0.200pt]{4.818pt}{0.400pt}}
\put(30.0,285){\rule[-0.200pt]{4.818pt}{0.400pt}}
\put(30.0,320){\rule[-0.200pt]{4.818pt}{0.400pt}}
\put(30.0,355){\rule[-0.200pt]{4.818pt}{0.400pt}}

\put(50,40){\usebox{\plotpoint}}
\scriptsize{
\put(-80,225){\makebox(0,0){height}}
\put(10,75){\makebox(0,0)[r]{0}}
\put(10,110){\makebox(0,0)[r]{1}}
\put(10,145){\makebox(0,0)[r]{2}}
\put(10,180){\makebox(0,0)[r]{3}}
\put(10,215){\makebox(0,0)[r]{4}}
\put(10,250){\makebox(0,0)[r]{5}}
\put(10,285){\makebox(0,0)[r]{6}}
\put(10,320){\makebox(0,0)[r]{7}}
\put(10,355){\makebox(0,0)[r]{8}}
\put(40,-15){\makebox(0,0)[l]{$1$}}
\put(80,-15){\makebox(0,0)[l]{$2$}}
\put(120,-15){\makebox(0,0)[l]{$3$}}
\put(150,-15){\makebox(0,0)[l]{$4$}}
\put(185,-15){\makebox(0,0)[l]{$5$}}
\put(220,-15){\makebox(0,0)[l]{$6$}}
\put(250,-15){\makebox(0,0)[l]{$\ldots$}}
\put(320,-10){\makebox(0,0)[l]{site number}}
\put(400,350){\makebox(0,0)[l]{Gas of tiles}}
\put(110,260){\makebox(0,0)[l]{A}}
\put(500,270){\makebox(0,0)[l]{B}}
\put(570,250){\makebox(0,0)[l]{C}}
\put(750,80){\makebox(0,0)[l]{Substrate}}
}
\put(125,150){\vector(0,-1){60}}
\put(510,220){\vector(0,-1){100}}
\put(740,80){\vector(-1,0){60}}
\put(580,180){\vector(0,-1){75}}
\Huge{\put(90,65){\raisebox{-.8pt}{\makebox(0,0){$\Diamond$}}}}
\Huge{\put(160,65){\raisebox{-.8pt}{\makebox(0,0){$\Diamond$}}}}
\Huge{\put(230,65){\raisebox{-.8pt}{\makebox(0,0){$\Diamond$}}}}
\Huge{\put(300,65){\raisebox{-.8pt}{\makebox(0,0){$\Diamond$}}}}
\Huge{\put(370,65){\raisebox{-.8pt}{\makebox(0,0){$\Diamond$}}}}
\Huge{\put(440,65){\raisebox{-.8pt}{\makebox(0,0){$\Diamond$}}}}
\Huge{\put(510,65){\raisebox{-.8pt}{\makebox(0,0){$\Diamond$}}}}
\Huge{\put(580,65){\raisebox{-.8pt}{\makebox(0,0){$\Diamond$}}}}
\Huge{\put(650,65){\raisebox{-.8pt}{\makebox(0,0){$\Diamond$}}}}

\Huge{\put(125,250){\raisebox{-.8pt}{\makebox(0,0){$\Diamond$}}}}
\Huge{\put(510,280){\raisebox{-.8pt}{\makebox(0,0){$\Diamond$}}}}
\Huge{\put(580,250){\raisebox{-.8pt}{\makebox(0,0){$\Diamond$}}}}

\Huge{\put(335,120){\raisebox{-.8pt}{\makebox(0,0){$\Diamond$}}}}
\Huge{\put(405,120){\raisebox{-.8pt}{\makebox(0,0){$\Diamond$}}}}
\Huge{\put(475,120){\raisebox{-.8pt}{\makebox(0,0){$\Diamond$}}}}
\Huge{\put(440,175){\raisebox{-.8pt}{\makebox(0,0){$\Diamond$}}}}

\put(50.0,410.0){\rule[-0.200pt]{118pt}{0.400pt}}
\end{picture} &
    \par\hspace{10mm} 
\setlength{\unitlength}{0.180900pt}
\ifx\plotpoint\undefined\newsavebox{\plotpoint}\fi
\sbox{\plotpoint}{\rule[-0.200pt]{0.400pt}{0.400pt}}%
\begin{picture}(750,450)(0,0)
\sbox{\plotpoint}{\rule[-0.200pt]{0.400pt}{0.400pt}}%
\put(50.0,40.0){\rule[-0.200pt]{0.400pt}{66pt}}
\put(700.0,40.0){\rule[-0.200pt]{0.400pt}{66pt}}
\put(30,75){\rule[-0.200pt]{4.818pt}{0.400pt}}
\put(30.0,110){\rule[-0.200pt]{4.818pt}{0.400pt}}
\put(30.0,145){\rule[-0.200pt]{4.818pt}{0.400pt}}
\put(30.0,180){\rule[-0.200pt]{4.818pt}{0.400pt}}
\put(30.0,215){\rule[-0.200pt]{4.818pt}{0.400pt}}
\put(30.0,250){\rule[-0.200pt]{4.818pt}{0.400pt}}
\put(30.0,285){\rule[-0.200pt]{4.818pt}{0.400pt}}
\put(30.0,320){\rule[-0.200pt]{4.818pt}{0.400pt}}
\put(30.0,355){\rule[-0.200pt]{4.818pt}{0.400pt}}

\put(50,40){\usebox{\plotpoint}}
\scriptsize{
\put(-80,225){\makebox(0,0){height}}
\put(10,75){\makebox(0,0)[r]{0}}
\put(10,110){\makebox(0,0)[r]{1}}
\put(10,145){\makebox(0,0)[r]{2}}
\put(10,180){\makebox(0,0)[r]{3}}
\put(10,215){\makebox(0,0)[r]{4}}
\put(10,250){\makebox(0,0)[r]{5}}
\put(10,285){\makebox(0,0)[r]{6}}
\put(10,320){\makebox(0,0)[r]{7}}
\put(10,355){\makebox(0,0)[r]{8}}
\put(40,-15){\makebox(0,0)[l]{$1$}}
\put(80,-15){\makebox(0,0)[l]{$2$}}
\put(120,-15){\makebox(0,0)[l]{$3$}}
\put(150,-15){\makebox(0,0)[l]{$4$}}
\put(185,-15){\makebox(0,0)[l]{$5$}}
\put(220,-15){\makebox(0,0)[l]{$6$}}
\put(250,-15){\makebox(0,0)[l]{$\ldots$}}
\put(320,-10){\makebox(0,0)[l]{site number}}
\put(100,350){\makebox(0,0)[l]{Gas of tiles}}
\put(110,120){\makebox(0,0)[l]{A}}
\put(500,270){\makebox(0,0)[l]{B}}
\put(570,250){\makebox(0,0)[l]{C}}
\put(750,80){\makebox(0,0)[l]{Substrate}}
}
\put(510,320){\vector(0,1){50}}
\put(740,80){\vector(-1,0){60}}
\put(580,300){\vector(0,1){75}}
\Huge{\put(90,65){\raisebox{-.8pt}{\makebox(0,0){$\Diamond$}}}}
\Huge{\put(160,65){\raisebox{-.8pt}{\makebox(0,0){$\Diamond$}}}}
\Huge{\put(230,65){\raisebox{-.8pt}{\makebox(0,0){$\Diamond$}}}}
\Huge{\put(300,65){\raisebox{-.8pt}{\makebox(0,0){$\Diamond$}}}}
\Huge{\put(370,65){\raisebox{-.8pt}{\makebox(0,0){$\Diamond$}}}}
\Huge{\put(440,65){\raisebox{-.8pt}{\makebox(0,0){$\Diamond$}}}}
\Huge{\put(510,65){\raisebox{-.8pt}{\makebox(0,0){$\Diamond$}}}}
\Huge{\put(580,65){\raisebox{-.8pt}{\makebox(0,0){$\Diamond$}}}}
\Huge{\put(650,65){\raisebox{-.8pt}{\makebox(0,0){$\Diamond$}}}}

\Huge{\put(125,120){\raisebox{-.8pt}{\makebox(0,0){$\Diamond$}}}}
\Huge{\put(510,270){\raisebox{-.8pt}{\makebox(0,0){$\Diamond$}}}}
\Huge{\put(580,250){\raisebox{-.8pt}{\makebox(0,0){$\Diamond$}}}}

\Huge{\put(335,120){\raisebox{-.8pt}{\makebox(0,0){$\Diamond$}}}}
\Huge{\put(405,170){\raisebox{-.8pt}{\makebox(0,0){$\Diamond$}}}}
\Huge{\put(475,170){\raisebox{-.8pt}{\makebox(0,0){$\Diamond$}}}}
\Huge{\put(440,225){\raisebox{-.8pt}{\makebox(0,0){$\Diamond$}}}}

\put(50.0,410.0){\rule[-0.200pt]{118pt}{0.400pt}}
\end{picture}\\
  \end{tabular}
  \caption{Three possible cases shown. 
    (A) tile attaches to the surface, 
    (B) tile removes a layer, 
    (C) tile is reflected.
  }
  \label{fig:cartoon}
\end{figure}

The Raise and Peel Model (RPM) describes a growing and fluctuating interface. An initial configuration is chosen and tiles are dropped onto the surface with a certain probability. Three different processes can happen  as shown in figure \ref{fig:cartoon}. 
With some probability $P_i = \frac{1}{L+a-1}$ a tile lands in site $i = 1,...,L-1$, while in RPMW the site i=0 is chosen with some probability
$P_0 = \frac{a}{L+a-1}$. Depending on the slope of surface at the $i^{th}$ site, one of three things can occur.
\begin{itemize}
\item Case A: \underline{Tile hits a local minimum}\\
  if site $i>0$, with some probability $u_a$ the tile attaches to the substrate, else if site i=0 is chosen then half-a-tile attaches to the boundary with rate 1.
\item Case B: \underline{Tile hits a slope}\\
  With Probability $u_b$ the tile peels off tiles within a cluster   such as the local local height at every site in the cluster decreases by two: $h_i \rightarrow h_i -2$ 
  (a tile has a height of 2); In other words, a tile may only remove \underline{one} layer of tiles above the point 
  of contact.
\item Case C: \underline{Tile hits a local maxima}\\
  Tile reflects and nothing happens.
\end{itemize}

This stochastic process in continuum time is given by the \emph{master equation} \cite{Alcaraz9,Golinelli}
\begin{equation}
 \frac{d}{dt} P_\alpha(t) = - \sum_\beta H_{\alpha,\beta}  P_\beta(t) 
\end{equation}
where $P_\alpha(t)$ is the (unnormalized) probability of finding the system in one of the states $| \alpha \rangle $ 
at time t, and $H_{\alpha,\beta}$ is the rate for the transition $| \alpha \rangle  \rightarrow | \beta \rangle $. 
Since this is an intensity matrix, there is at least one zero eigenvalue \cite{alcaraz} and its corresponding eigenvector $|0 \rangle$ gives the probabilities in the \emph{stationary} state

\be 
\langle 0 | H = 0,  ~~~ \langle 0 |  = (1,1,\ldots,1) 
\ee
\be
H|0 \rangle = 0,~~~ |0 \rangle = \sum_\alpha P_\alpha | \alpha \rangle,~~~~~~~~P_\alpha = \lim_{t\to \infty} P_\alpha(t) 
\ee 
For the  special case in which the rates are equal $u_a = u_d$, the
Hamiltonians can be written in terms of 
the Temperley-Lieb algebra defined in terms of generators $e_1,e_2,\dots,e_L$ satisfying the following commutation
relations \cite{Pearce2}
\begin{align}\label{eqn:TLalbegra}
  e_i^2 = e_i \nonumber \\
  e_i e_{i+1} e_i = e_i \nonumber \\
  e_i e_{i-1} e_i = e_i \nonumber \\
  [e_i,e_j] = 0~~~for~~|i-j| \geq 2 
\end{align}
while the one-boundary term $e_0$ at site $i=0$ satisfies the following constraint \cite{Nichols,gier3}
\begin{align}
   e_i^0 = e_0, \hspace{20mm}
   e_1 e_{0} e_1 = e_1, \hspace{20mm}
   e_{0} e_i = e_i e_0  ~~~if~~i > 1
\end{align}
  
There are many representation to the Temperley-Lieb algebra including ``blob-algebra''.
For our purposes it becomes convenient to view the generators in terms of a tile
and half-tile at the boundary.

\be
e_0 = 
 \put(11,11){\line(1,-1){10}}
 \put(11,-9){\line(1,1){10}}
 \put(11,-9){\line(0,1){20}}
 \multiput(9.5,-12)(0,10){1}{\tiny \vdots} \multiput(9.5,0)(0,10){1}{\tiny \vdots}  \multiput(9.5,4)(0,10){1}{\tiny \vdots}
\put(13,-15){\makebox(0,0){0}}
~~~~~~~~~~~~~~e_i = 
 \put(1,1){\line(1,1){10}}
 \put(11,11){\line(1,-1){10}}
 \put(1,1){\line(1,-1){10}}
 \put(11,-9){\line(1,1){10}}
 \multiput(10,-12)(0,10){1}{\tiny \vdots} \multiput(10,0)(0,10){1}{\tiny \vdots}  \multiput(10,4)(0,10){1}{\tiny \vdots}
\put(13,-15){\makebox(0,0){i}}
\ee

Products of generators at different sites, following the algebra relations, reduce the products 
to a subset of unique configurations. This is illustrated by the following two examples.
\be
e_4 e_6 e_1 e_3 e_5 e_7 =  
\setlength{\unitlength}{1.0pt}
 \put(1,1){\line(1,1){10}}
 \put(11,11){\line(1,-1){10}}
 \put(1,1){\line(1,-1){10}}
 \put(11,-9){\line(1,1){10}}
 \put(21,1){\line(1,1){10}}
 \put(31,11){\line(1,-1){10}}
 \put(21,1){\line(1,-1){10}}
 \put(31,-9){\line(1,1){10}}
 \put(41,1){\line(1,1){10}}
 \put(51,11){\line(1,-1){10}}
 \put(41,1){\line(1,-1){10}}
 \put(51,-9){\line(1,1){10}}
 \put(61,1){\line(1,1){10}}
 \put(71,11){\line(1,-1){10}}
 \put(61,1){\line(1,-1){10}}
 \put(71,-9){\line(1,1){10}}
 \put(31,11){\line(1,1){10}}
 \put(41,21){\line(1,-1){10}}
 \put(51,11){\line(1,1){10}}
 \put(61,21){\line(1,-1){10}}
 \multiput(0,-12)(0,10){1}{\tiny \vdots}
 \multiput(0,0)(0,10){1}{\tiny \vdots}  
 \multiput(0,4)(0,10){1}{\tiny \vdots}
 \multiput(0,8)(0,10){1}{\tiny \vdots}
 \multiput(0,12)(0,10){1}{\tiny \vdots}
 \multiput(10,-12)(0,10){1}{\tiny \vdots}
 \multiput(10,0)(0,10){1}{\tiny \vdots}  
 \multiput(10,4)(0,10){1}{\tiny \vdots}
 \multiput(10,8)(0,10){1}{\tiny \vdots}
 \multiput(10,12)(0,10){1}{\tiny \vdots}
 \multiput(20,-12)(0,10){1}{\tiny \vdots}
 \multiput(20,0)(0,10){1}{\tiny \vdots}  
 \multiput(20,4)(0,10){1}{\tiny \vdots}
 \multiput(20,8)(0,10){1}{\tiny \vdots}
 \multiput(20,12)(0,10){1}{\tiny \vdots}
 \multiput(30,-12)(0,10){1}{\tiny \vdots}
 \multiput(30,0)(0,10){1}{\tiny \vdots}  
 \multiput(30,4)(0,10){1}{\tiny \vdots}
 \multiput(30,8)(0,10){1}{\tiny \vdots}
 \multiput(30,12)(0,10){1}{\tiny \vdots}
 \multiput(40,-12)(0,10){1}{\tiny \vdots}
 \multiput(40,0)(0,10){1}{\tiny \vdots}  
 \multiput(40,4)(0,10){1}{\tiny \vdots}
 \multiput(40,8)(0,10){1}{\tiny \vdots}
 \multiput(40,12)(0,10){1}{\tiny \vdots}
 \multiput(50,-12)(0,10){1}{\tiny \vdots}
 \multiput(50,0)(0,10){1}{\tiny \vdots}  
 \multiput(50,4)(0,10){1}{\tiny \vdots}
 \multiput(50,8)(0,10){1}{\tiny \vdots}
 \multiput(50,12)(0,10){1}{\tiny \vdots}
 \multiput(60,-12)(0,10){1}{\tiny \vdots}
 \multiput(60,0)(0,10){1}{\tiny \vdots}  
 \multiput(60,4)(0,10){1}{\tiny \vdots}
 \multiput(60,8)(0,10){1}{\tiny \vdots}
 \multiput(60,12)(0,10){1}{\tiny \vdots}
 \multiput(70,-12)(0,10){1}{\tiny \vdots}
 \multiput(70,0)(0,10){1}{\tiny \vdots}  
 \multiput(70,4)(0,10){1}{\tiny \vdots}
 \multiput(70,8)(0,10){1}{\tiny \vdots}
 \multiput(70,12)(0,10){1}{\tiny \vdots}
 \multiput(80,-12)(0,10){1}{\tiny \vdots}
 \multiput(80,0)(0,10){1}{\tiny \vdots}  
 \multiput(80,4)(0,10){1}{\tiny \vdots}
 \multiput(80,8)(0,10){1}{\tiny \vdots}
 \multiput(80,12)(0,10){1}{\tiny \vdots}
\put(1,-20){\makebox(0,0){0}}
\put(11,-20){\makebox(0,0){1}}
\put(21,-20){\makebox(0,0){2}}
\put(31,-20){\makebox(0,0){3}}
\put(41,-20){\makebox(0,0){4}}
\put(51,-20){\makebox(0,0){5}}
\put(61,-20){\makebox(0,0){6}}
\put(71,-20){\makebox(0,0){7}}
\put(81,-20){\makebox(0,0){8}}
\ee

\be
e_0 e_1 e_3 e_0 e_2 e_4 e_6 e_1 e_3 e_5 e_7 =  
 \put(1,1){\line(1,1){10}}
 \put(11,11){\line(1,-1){10}}
 \put(1,1){\line(1,-1){10}}
 \put(11,-9){\line(1,1){10}}
 \put(21,1){\line(1,1){10}}
 \put(31,11){\line(1,-1){10}}
 \put(21,1){\line(1,-1){10}}
 \put(31,-9){\line(1,1){10}}
 \put(41,1){\line(1,1){10}}
 \put(51,11){\line(1,-1){10}}
 \put(41,1){\line(1,-1){10}}
 \put(51,-9){\line(1,1){10}}
 \put(61,1){\line(1,1){10}}
 \put(71,11){\line(1,-1){10}}
 \put(61,1){\line(1,-1){10}}
 \put(71,-9){\line(1,1){10}}
 \put(11,11){\line(1,1){10}}
 \put(21,21){\line(1,-1){10}}
 \put(31,11){\line(1,1){10}}
 \put(41,21){\line(1,-1){10}}
 \put(51,11){\line(1,1){10}}
 \put(61,21){\line(1,-1){10}}
 \put(21,21){\line(1,1){10}}
 \put(31,31){\line(1,-1){10}}
 \put(1,21){\line(1,1){10}}
 \put(11,31){\line(1,-1){10}}
 \put(1,21){\line(1,-1){10}}
 \put(1,41){\line(1,-1){10}}
 \put(1,21){\line(0,1){20}}
 \put(1,21){\line(0,-1){20}}
 \multiput(0,-12)(0,10){1}{\tiny \vdots}
 \multiput(0,0)(0,10){1}{\tiny \vdots}  
 \multiput(0,4)(0,10){1}{\tiny \vdots}
 \multiput(0,8)(0,10){1}{\tiny \vdots}
 \multiput(0,12)(0,10){1}{\tiny \vdots}
 \multiput(10,-12)(0,10){1}{\tiny \vdots}
 \multiput(10,0)(0,10){1}{\tiny \vdots}  
 \multiput(10,4)(0,10){1}{\tiny \vdots}
 \multiput(10,8)(0,10){1}{\tiny \vdots}
 \multiput(10,12)(0,10){1}{\tiny \vdots}
 \multiput(20,-12)(0,10){1}{\tiny \vdots}
 \multiput(20,0)(0,10){1}{\tiny \vdots}  
 \multiput(20,4)(0,10){1}{\tiny \vdots}
 \multiput(20,8)(0,10){1}{\tiny \vdots}
 \multiput(20,12)(0,10){1}{\tiny \vdots}
 \multiput(30,-12)(0,10){1}{\tiny \vdots}
 \multiput(30,0)(0,10){1}{\tiny \vdots}  
 \multiput(30,4)(0,10){1}{\tiny \vdots}
 \multiput(30,8)(0,10){1}{\tiny \vdots}
 \multiput(30,12)(0,10){1}{\tiny \vdots}
 \multiput(40,-12)(0,10){1}{\tiny \vdots}
 \multiput(40,0)(0,10){1}{\tiny \vdots}  
 \multiput(40,4)(0,10){1}{\tiny \vdots}
 \multiput(40,8)(0,10){1}{\tiny \vdots}
 \multiput(40,12)(0,10){1}{\tiny \vdots}
 \multiput(50,-12)(0,10){1}{\tiny \vdots}
 \multiput(50,0)(0,10){1}{\tiny \vdots}  
 \multiput(50,4)(0,10){1}{\tiny \vdots}
 \multiput(50,8)(0,10){1}{\tiny \vdots}
 \multiput(50,12)(0,10){1}{\tiny \vdots}
 \multiput(60,-12)(0,10){1}{\tiny \vdots}
 \multiput(60,0)(0,10){1}{\tiny \vdots}  
 \multiput(60,4)(0,10){1}{\tiny \vdots}
 \multiput(60,8)(0,10){1}{\tiny \vdots}
 \multiput(60,12)(0,10){1}{\tiny \vdots}
 \multiput(70,-12)(0,10){1}{\tiny \vdots}
 \multiput(70,0)(0,10){1}{\tiny \vdots}  
 \multiput(70,4)(0,10){1}{\tiny \vdots}
 \multiput(70,8)(0,10){1}{\tiny \vdots}
 \multiput(70,12)(0,10){1}{\tiny \vdots}
 \multiput(80,-12)(0,10){1}{\tiny \vdots}
 \multiput(80,0)(0,10){1}{\tiny \vdots}  
 \multiput(80,4)(0,10){1}{\tiny \vdots}
 \multiput(80,8)(0,10){1}{\tiny \vdots}
 \multiput(80,12)(0,10){1}{\tiny \vdots}
\put(1,-20){\makebox(0,0){0}}
\put(11,-20){\makebox(0,0){1}}
\put(21,-20){\makebox(0,0){2}}
\put(31,-20){\makebox(0,0){3}}
\put(41,-20){\makebox(0,0){4}}
\put(51,-20){\makebox(0,0){5}}
\put(61,-20){\makebox(0,0){6}}
\put(71,-20){\makebox(0,0){7}}
\put(81,-20){\makebox(0,0){8}}
\ee

In terms of these generators, the stochastic Hamiltonians H for RPM and $H^{(a)}$ for RPMW with a rate $ a$ 
at the boundary can be written as:
\begin{equation}\label{eqn:rpmsH}
 H = \sum_{i=1}^{L-1} (1-e_i) ~~~~~~~~~  H^{(a)} = a(1-e_0) + \sum_{i=1}^{L-1} (1-e_i) 
\end{equation}

The ground state eigenvectors of the intensity matrices (Eqn. \ref{eqn:rpmsH}) have remarkable combinatorial properties
\cite{alcaraz,alcaraz_facets,gier2}. The normalization on the ground state eigenvectors will be used in section \ref{sec:conj} 
to derive conjectures for  probabilities.
The following two examples illustrate the combinatorial properties for a small lattice with L=6 in RPM and for L=4 in RPMW.\\
\par\hspace{30mm}\rrra \rrrb \rrrc \rrrd \rrre \hspace{18mm}\wwwf \wwwb \wwwd~~ \wwwe~~ \wwwa~~ \wwwc 
\[
H = 
\left( {\begin{array}{ccccc}
    -2 & ~2 & ~2 & ~0 & ~2\\
    ~1 & -3 & ~0 & ~1 & ~0 \\
    ~1 & ~0 & -3 & ~1 & ~0 \\
    ~0 & ~1 & ~1 & -3 & ~2 \\
    ~0 & ~0 & ~0 & ~1 & -4 \\
\end{array} } \right)\qquad
H^{(a)} =
\left( {\begin{array}{cccccc}
    -3 & ~0 & ~0 & ~a & ~0 & ~0\\
    ~1 & -2 & ~1 & ~0 & ~a & ~0\\
    ~1 & ~1 & -2 & ~1 & ~0 & ~a\\
    ~1 & ~0 & ~1 &-2-a& ~0 & ~0\\
    ~0 & ~1 & ~0 & ~1 &-1-a& ~2\\
    ~0 & ~0 & ~0 & ~0 & ~1 &-2-a\\
\end{array} } \right)
\]

The wave functions $|0\rangle$  ($|0\rangle^{(a)}$) normalized to have the smallest entry equal to 1 (or a) are given by:
\begin{equation}\label{eqn:statst}
 |0 \rangle = (11,5,5,4,1) \hspace{20mm} |0\rangle^{(a)} = (a^2,3a(2+a),2a(3+a),3a,3(2+a),3)
\end{equation}
The normalization factor  $\langle 0| 0 \rangle$ for RPM is then given by (L=2n)
\begin{equation}\label{eqn:rpmN}
\langle 0| 0 \rangle = A_V (2n + 1) = 1, 3, 26, 646, . . .
\end{equation}
while the normalization factors  $\langle 0| 0 \rangle$ for RPMW with a=1 is given by
\begin{equation}\label{eqn:rpmwN}
^{(1)}\langle 0| 0 \rangle^{(1)} = N_8(2n) A_V (2n + 1) = 2, 33, 26, 4420, . . .
\end{equation}
where $A_V (2n + 1)$ is the number of vertically symmetric $(2n+1)\times(2n+1)$ 
alternating sign matrices \cite{mitra,gier5,Propp}
\begin{equation}
 A_V (2n + 1) = \Pi_{j=0}^{n-1} (3j+2)\frac{(2j+1)!(6j+3)!}{(4j+2)!(4j+3)!}=1,3,26,646,\cdots
\end{equation}
and $N_8(2n)$ is the number of cyclically symmetric transpose complement plane partitions \cite{mitra,gier3} 
\begin{equation}
 N_8(2n) = \Pi_{j=0}^{n-1} (3j+1)\frac{(2j)!(6j)!}{(4j)!(4j+1)!}=1,2,11,170,7429,\cdots
\end{equation}

\section{Energy Spectra and space-time phenomena}\label{sec:spectra}
For $u \equiv u_a/u_d=1$, the finite-size corrections to the energy spectra
of the intensity matrices $H$  and $H^{(a)}$ are given by a conformal field
theory with central charge ($c=0$) \cite{Pearce,magic,nonlocalSS}
\begin{equation}
  E_n = E_0 + \frac{\pi v (\Delta_s+ k)}{L} + O(L^{-1})~~~~~~~k \in \mathbb{Z}~and~v = \frac{3 \sqrt{3}}{2}
\end{equation}
\begin{equation}
  \Delta_{s} = \frac{s(2s-1)}{3} = 0,0,\frac{1}{3},1 \cdots ~~~~ s = 0,\frac{1}{2},1,\frac{3}{2},\cdots
\end{equation}
In table~\ref{table:spectra} the excited energy states $E_n$ for different spin sectors 
are compared to the numerical estimations 
obtained by diagonalizing the intensity matrices $H$ and $H^{(a)}$  for RPM and RPMW (Eqn. \ref{eqn:rpmsH}).

\begin{table}[h]
  \caption{Excited energy states in units of $\frac{v \pi}{L}$ for different spin sectors, and numerical
    approximations to $E_n$ using L=18 for RPM  and L=16 RPMW}
  \centering
  \begin{tabular}{|c|cccccc|}
    \hline
    $\Delta_s + k $ & 0 &1 & 2 & 3 & 4 &5\\
    \hline \hline
    s = 0 & 0 & 2 & 3 & 4 & 4 & 5 \\
    $s = \frac{1}{2}$ & 0 & 1 & 2 & 3 & 3 & 4 \\
    $s = 1 \bigoplus s=1/2$ & 0 & 1 & 2 & 2 & 3 & 3 \\
    \hline \hline                                 
    $n $ & 0 &1 & 2 & 3 & 4 &5\\                                                                                                                           
    \hline \hline
    RPM & 0.0000 & 2.0009 & 3.0035 & 4.0247  &4.0257 & 5.037 \\
    RPMW & 0.0000 &  1.0015 & 2.0087 & 1.9954 & 3.0030 & 3.0158 \\
    \hline
  \end{tabular}
  \label{table:spectra}
\end{table}
The functional dependence of the time evolution of distributions can be predicted. The 
expectation value of observables can be described by stochastic dynamics as:
\begin{equation}\label{eqn:stochdyn}
  \langle X \rangle(t) = \langle 0 | X e^{-Ht} | \Psi(0) \rangle   
\end{equation}
where the initial state $| \Psi(0) \rangle$ 
can be expanded in a complete eigenbasis characterizing the system: $|\Psi(0)\rangle = \sum_n c_n | \psi_n \rangle$,
and H is the stochastic matrix describing the system. Since H is an intensity matrix, the lowest eigenvalue is zero, hence the lowest non-zero eigenvalue $E_1$ is expected to dominate the time evolution for large times.

\begin{figure}[h]
  \centering
  \includegraphics[scale=.85]{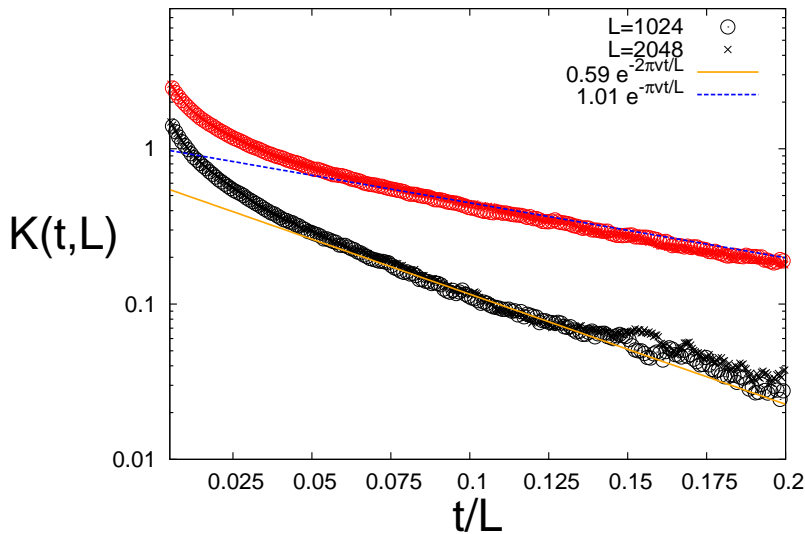}
  \caption{The time evolution of the average number of clusters K(t) for RPM and RPMW. 
    The lines are the expected decays with $K(t,L) \propto e^{-E_1t/L}$ with $E_1 = \frac{2 v \pi}{L}$  (RPM) 
    and $E_1 = \frac{ v \pi}{L}$ (RPMW) }
  \label{fig:Kt}
\end{figure}

Let us consider the effect of adding a wall on temporal profiles of quantities describing the system. 
The temporal average of a quantity will be denoted by the Family-Vicsek \cite{Family} scaling form:
\begin{equation}\label{eqn:FV}
  X(t,L) = \frac{x(t,L)}{x(L)}  - 1  \sim X(\frac{t}{L^z})
\end{equation}
The average number of clusters $ K(x) = \langle \sum_j^L \delta_{h_i,0} \rangle$ is plotted in Fig.~\ref{fig:Kt} in the
form given by Eqn.(\ref{eqn:FV}). The data collapse shows that the average number of clusters $K(t,L)$ 
has a critical exponent given by $z=1$ for different values of the rate a: (a=0 RPM, a=1 RPMW). 
The long time decay is described well by the exponential given by $E_1 = \frac{2 \pi v}{L}$ for RPM 
and by $E_1 = \frac{\pi v}{L}$ for RPMW as is expected from Eqn.(\ref{eqn:stochdyn}) for large times.

\section{Avalanches}\label{sec:avalanches}

\begin{figure}[htpb]
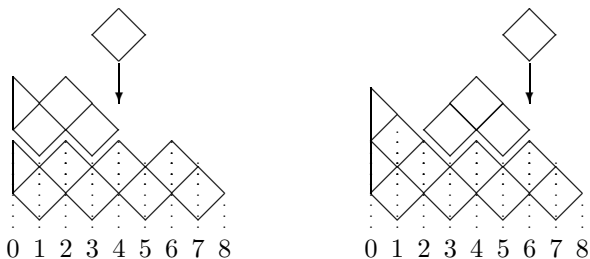

\begin{tabular}{cc}
 $
 \put(41,50){\vector(0,-1){15}}
 \put(31,61){\line(1,1){10}}
 \put(41,71){\line(1,-1){10}}
 \put(31,61){\line(1,-1){10}}
 \put(41,51){\line(1,1){10}}
 \put(1,1){\line(1,1){10}}
 \put(11,11){\line(1,-1){10}}
 \put(1,1){\line(1,-1){10}}
 \put(11,-9){\line(1,1){10}}
 \put(21,1){\line(1,1){10}}
 \put(31,11){\line(1,-1){10}}
 \put(21,1){\line(1,-1){10}}
 \put(31,-9){\line(1,1){10}}
 \put(41,1){\line(1,1){10}}
 \put(51,11){\line(1,-1){10}}
 \put(41,1){\line(1,-1){10}}
 \put(51,-9){\line(1,1){10}}
 \put(61,1){\line(1,1){10}}
 \put(71,11){\line(1,-1){10}}
 \put(61,1){\line(1,-1){10}}
 \put(71,-9){\line(1,1){10}}
 \put(11,11){\line(1,1){10}}
 \put(21,21){\line(1,-1){10}}
 \put(31,11){\line(1,1){10}}
 \put(41,21){\line(1,-1){10}}
 \put(51,11){\line(1,1){10}}
 \put(61,21){\line(1,-1){10}}
 \put(21,25){\line(1,1){10}}
 \put(31,35){\line(1,-1){10}}
 \put(11,35){\line(1,1){10}}
 \put(21,45){\line(1,-1){10}}
 \put(1,25){\line(1,1){10}}
 \put(11,35){\line(1,-1){10}}
 \put(1,25){\line(1,-1){10}}
 \put(1,45){\line(1,-1){10}}
 \put(1,25){\line(0,1){20}}
 \put(11,15){\line(1,1){10}}
 \put(21,25){\line(1,-1){10}}
 \put(31,15){\line(1,1){10}}
 \put(1,21){\line(1,-1){10}}
 \put(1,21){\line(0,-1){20}}
 \multiput(0,-12)(0,10){1}{\tiny \vdots}
 \multiput(0,0)(0,10){1}{\tiny \vdots}  
 \multiput(0,4)(0,10){1}{\tiny \vdots}
 \multiput(0,8)(0,10){1}{\tiny \vdots}
 \multiput(0,12)(0,10){1}{\tiny \vdots}
 \multiput(10,-12)(0,10){1}{\tiny \vdots}
 \multiput(10,0)(0,10){1}{\tiny \vdots}  
 \multiput(10,4)(0,10){1}{\tiny \vdots}
 \multiput(20,-12)(0,10){1}{\tiny \vdots}
 \multiput(20,0)(0,10){1}{\tiny \vdots}  
 \multiput(20,4)(0,10){1}{\tiny \vdots}
 \multiput(20,8)(0,10){1}{\tiny \vdots}
 \multiput(20,12)(0,10){1}{\tiny \vdots}
 \multiput(30,-12)(0,10){1}{\tiny \vdots}
 \multiput(30,0)(0,10){1}{\tiny \vdots}  
 \multiput(30,4)(0,10){1}{\tiny \vdots}
 \multiput(40,-12)(0,10){1}{\tiny \vdots}
 \multiput(40,0)(0,10){1}{\tiny \vdots}  
 \multiput(40,4)(0,10){1}{\tiny \vdots}
 \multiput(40,8)(0,10){1}{\tiny \vdots}
 \multiput(40,12)(0,10){1}{\tiny \vdots}
 \multiput(50,-12)(0,10){1}{\tiny \vdots}
 \multiput(50,0)(0,10){1}{\tiny \vdots}  
 \multiput(50,4)(0,10){1}{\tiny \vdots}
 \multiput(60,-12)(0,10){1}{\tiny \vdots}
 \multiput(60,0)(0,10){1}{\tiny \vdots}  
 \multiput(60,4)(0,10){1}{\tiny \vdots}
 \multiput(60,8)(0,10){1}{\tiny \vdots}
 \multiput(60,12)(0,10){1}{\tiny \vdots}
 \multiput(70,-12)(0,10){1}{\tiny \vdots}
 \multiput(70,0)(0,10){1}{\tiny \vdots}  
 \multiput(70,4)(0,10){1}{\tiny \vdots}
 \multiput(80,-12)(0,10){1}{\tiny \vdots}
\put(1,-20){\makebox(0,0){0}}
\put(11,-20){\makebox(0,0){1}}
\put(21,-20){\makebox(0,0){2}}
\put(31,-20){\makebox(0,0){3}}
\put(41,-20){\makebox(0,0){4}}
\put(51,-20){\makebox(0,0){5}}
\put(61,-20){\makebox(0,0){6}}
\put(71,-20){\makebox(0,0){7}}
\put(81,-20){\makebox(0,0){8}}
$ & 
\hspace{45mm}
$
 \put(61,50){\vector(0,-1){15}}
 \put(51,61){\line(1,1){10}}
 \put(61,71){\line(1,-1){10}}
 \put(51,61){\line(1,-1){10}}
 \put(61,51){\line(1,1){10}}
 \put(1,1){\line(1,1){10}}
 \put(11,11){\line(1,-1){10}}
 \put(1,1){\line(1,-1){10}}
 \put(11,-9){\line(1,1){10}}
 \put(21,1){\line(1,1){10}}
 \put(31,11){\line(1,-1){10}}
 \put(21,1){\line(1,-1){10}}
 \put(31,-9){\line(1,1){10}}
 \put(41,1){\line(1,1){10}}
 \put(51,11){\line(1,-1){10}}
 \put(41,1){\line(1,-1){10}}
 \put(51,-9){\line(1,1){10}}
 \put(61,1){\line(1,1){10}}
 \put(71,11){\line(1,-1){10}}
 \put(61,1){\line(1,-1){10}}
 \put(71,-9){\line(1,1){10}}
 \put(11,11){\line(1,1){10}}
 \put(21,21){\line(1,-1){10}}
 \put(31,11){\line(1,1){10}}
 \put(41,21){\line(1,-1){10}}
 \put(51,11){\line(1,1){10}}
 \put(61,21){\line(1,-1){10}}
 \put(1,21){\line(1,1){10}}
 \put(11,31){\line(1,-1){10}}
 \put(1,21){\line(1,-1){10}}
 \put(1,41){\line(1,-1){10}}
 \put(1,21){\line(0,1){20}}
 \put(1,21){\line(0,-1){20}}
 \put(41,25){\line(1,-1){10}}
 \put(51,15){\line(1,1){10}}
 \put(41,25){\line(1,1){10}}
 \put(51,35){\line(1,-1){10}}
 \put(31,35){\line(1,-1){10}}
 \put(41,25){\line(1,1){10}}
 \put(31,35){\line(1,1){10}}
 \put(41,45){\line(1,-1){10}}
 \put(21,25){\line(1,-1){10}}
 \put(31,15){\line(1,1){10}}
 \put(21,25){\line(1,1){10}}
 \put(31,35){\line(1,-1){10}}
 \multiput(0,-12)(0,10){1}{\tiny \vdots}
 \multiput(0,0)(0,10){1}{\tiny \vdots}  
 \multiput(0,4)(0,10){1}{\tiny \vdots}
 \multiput(0,8)(0,10){1}{\tiny \vdots}
 \multiput(0,12)(0,10){1}{\tiny \vdots}
 \multiput(10,-12)(0,10){1}{\tiny \vdots}
 \multiput(10,0)(0,10){1}{\tiny \vdots}  
 \multiput(10,4)(0,10){1}{\tiny \vdots}
 \multiput(10,8)(0,10){1}{\tiny \vdots}
 \multiput(10,12)(0,10){1}{\tiny \vdots}
 \multiput(10,16)(0,10){1}{\tiny \vdots}
 \multiput(20,-12)(0,10){1}{\tiny \vdots}
 \multiput(20,0)(0,10){1}{\tiny \vdots}  
 \multiput(20,4)(0,10){1}{\tiny \vdots}
 \multiput(20,8)(0,10){1}{\tiny \vdots}
 \multiput(20,12)(0,10){1}{\tiny \vdots}
 \multiput(30,-12)(0,10){1}{\tiny \vdots}
 \multiput(30,0)(0,10){1}{\tiny \vdots}  
 \multiput(30,4)(0,10){1}{\tiny \vdots}
 \multiput(40,-12)(0,10){1}{\tiny \vdots}
 \multiput(40,0)(0,10){1}{\tiny \vdots}  
 \multiput(40,4)(0,10){1}{\tiny \vdots}
 \multiput(40,8)(0,10){1}{\tiny \vdots}
 \multiput(40,12)(0,10){1}{\tiny \vdots}
 \multiput(50,-12)(0,10){1}{\tiny \vdots}
 \multiput(50,0)(0,10){1}{\tiny \vdots}  
 \multiput(50,4)(0,10){1}{\tiny \vdots}
 \multiput(60,-12)(0,10){1}{\tiny \vdots}
 \multiput(60,0)(0,10){1}{\tiny \vdots}  
 \multiput(60,4)(0,10){1}{\tiny \vdots}
 \multiput(60,8)(0,10){1}{\tiny \vdots}
 \multiput(60,12)(0,10){1}{\tiny \vdots}
 \multiput(70,-12)(0,10){1}{\tiny \vdots}
 \multiput(70,0)(0,10){1}{\tiny \vdots}  
 \multiput(70,4)(0,10){1}{\tiny \vdots}
 \multiput(80,-12)(0,10){1}{\tiny \vdots}
\put(1,-20){\makebox(0,0){0}}
\put(11,-20){\makebox(0,0){1}}
\put(21,-20){\makebox(0,0){2}}
\put(31,-20){\makebox(0,0){3}}
\put(41,-20){\makebox(0,0){4}}
\put(51,-20){\makebox(0,0){5}}
\put(61,-20){\makebox(0,0){6}}
\put(71,-20){\makebox(0,0){7}}
\put(81,-20){\makebox(0,0){8}}
$   
  \end{tabular}
  \centering
  \caption{Example of an even and odd avalanche in RPMW. Avalanches occurring on RPM release only an odd number of tiles}
  \label{fig:Avalanche}
\end{figure}

The raise and peel model exhibits events where layers are evaporated from the substrate when a tile from 
the gas hits the interface. The number of tiles removed defines the size of an avalanche. While this number is always
an odd number in RPM, in RPWM there is the possibility of an even number of tiles removed whenever an avalanche 
touches the boundary. This is illustrated in Figure \ref{fig:Avalanche}.

It is known that the raise and peel model exhibits self-organized criticality \cite{bak,dhar} in the regime for $u \geq 1$ \cite{alcaraz}.
Desorption processes being non--local results in avalanches lacking a
characteristic length-scale.  Their distribution $S(v,L)$ therefore appears as a power-law which might be described in the finite-size scaling (FSS) form \cite{Tebaldi,Alcaraz0}
\be\label{eqn:FSS}
 S(v,L) = v^{-\tau} F(\frac{v}{L^D})
\ee

In order to obtain the exponents, the method of moments is used \cite{alcaraz,Tebaldi}. Using the
scaling form (Eqn. \ref{eqn:FSS}) we have:
 \begin{align}
    \langle v^m \rangle_L & = \int S(v,L) v^m dv \\
    & = \int v^{-\tau} F(\frac{v}{L^D}) v^m dv \\
    & = \int w^{-\tau} L^{-D\tau} F(w) w^m w^{mD} L^D dw \\
    & =  L^{D(1+m-\tau)} \underbrace{\int  w^{m-\tau} F(w) dw} \\
    &   ~~~~~~~~~~~~~~~~~~~~~~~~~~~~ \Gamma_m 
  \end{align}
where we have used $w \equiv v/L^D$ to get the scaling dependence with L. 
We can get an estimate for the exponent by looking at the ratio:

\be 
  \langle v^m \rangle_L / \langle v^m \rangle_{L'}  = (L/L')^{\sigma(m)} 
\ee
and in this manner the exponent $\sigma(m)$ can be estimated as \cite{gier}:

\be\label{eqn:sigma}
  \sigma(m) = \frac{ln( \langle v^m \rangle_L / \langle v^m \rangle_{L'})}{ln(L/L')}   =
    \left\{ \begin{array}{lcl}
    0 & \mbox{for} & m < \tau - 1 \\ 
    D(1+m-\tau) & \mbox{for} & m > \tau - 1  
  \end{array}\right.
\ee

A linear fit to Eqn.(\ref{eqn:sigma}) for  $m > \tau - 1$ gives an estimate
for the values of $D$ and $\tau$. 
To get an idea of the spread of these values we ran several Monte-Carlo simulations to find the variation of the distribution
resulting from different seeds.  The results are shown in table \ref{table:exponents}.

\begin{table}[t]
    \caption{Estimates for the critical exponents in $S(v,L) \sim v^{-\tau} F(v/L^D)$ for even and odd avalanches using lattices L=4096 and L'=8192}
    \centering
    \begin{tabular}{c | c c l  l  l  }
       &  1/u &   & 1.0 & 0.45 & 0.005 \\
      \hline
       &RPM \cite{alcaraz}   &odd &1.004 & 1.026 & 1.006\\
      D &RPM[{\small this work}]&odd &$0.992 \pm 0.058$ & $1.015 \pm 0.006$ & $1.006 \pm 0.0001$\\
       &RPMW[{\small this work}]&odd &$0.994 \pm 0.025$ & $1.008 \pm 0.002$  & $1.006 \pm 0.0001$\\
       &RPMW[{\small this work}]&even &$0.980 \pm 0.117$ & $1.013 \pm 0.008$ & $1.006 \pm 0.0001$\\
      \hline
      &RPM \cite{alcaraz}         &odd &3.000 &2.25  & 2.00\\
      $\tau$ &RPM[{\small this work}]&odd &$2.977 \pm 0.079$ & $2.224 \pm 0.016$  & $2.011 \pm 0.001$ \\
      &RPMW[{\small this work}]      &odd &$3.003 \pm 0.071$ &$2.237 \pm 0.012$  & $2.011 \pm 0.001$ \\
      &RPMW[{\small this work}]      &even&$1.932 \pm 0.222$ &$1.280 \pm 0.074$  & $1.020 \pm 0.011$ \\
    \end{tabular}
    \label{table:exponents}
  \end{table}

\begin{figure}[b]
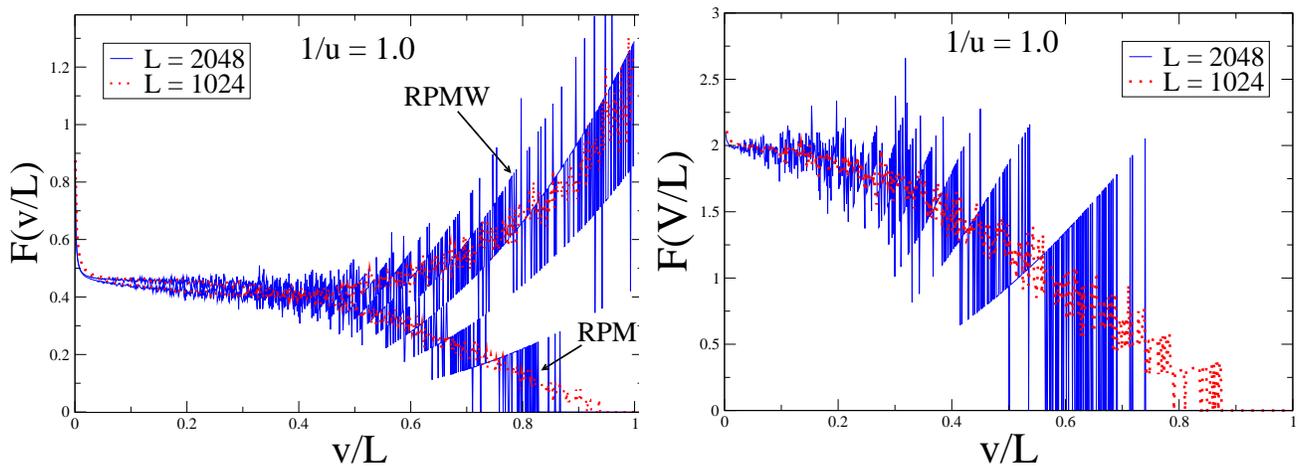

  \par\vspace{3mm}
  \begin{tabular}{cc}
    \includegraphics[scale=.35]{FIGS/Fv_u1.eps} &
    \includegraphics[scale=.35]{FIGS/Fve_u1.eps} \\
  \end{tabular}
  \caption{The scaling function $F(\frac{v}{L})$ with rates $1/u=1.0$ for $v \in$ {\it odd} number with {\bf $\tau=3.0$} and {\bf $D=1.0$} (LEFT) and $v \in$ {\it even} number with {\bf $\tau=2.0$} and {\bf $D=1.0$}}
  \label{fig:FVu1}
\end{figure}

\begin{figure}[ht]
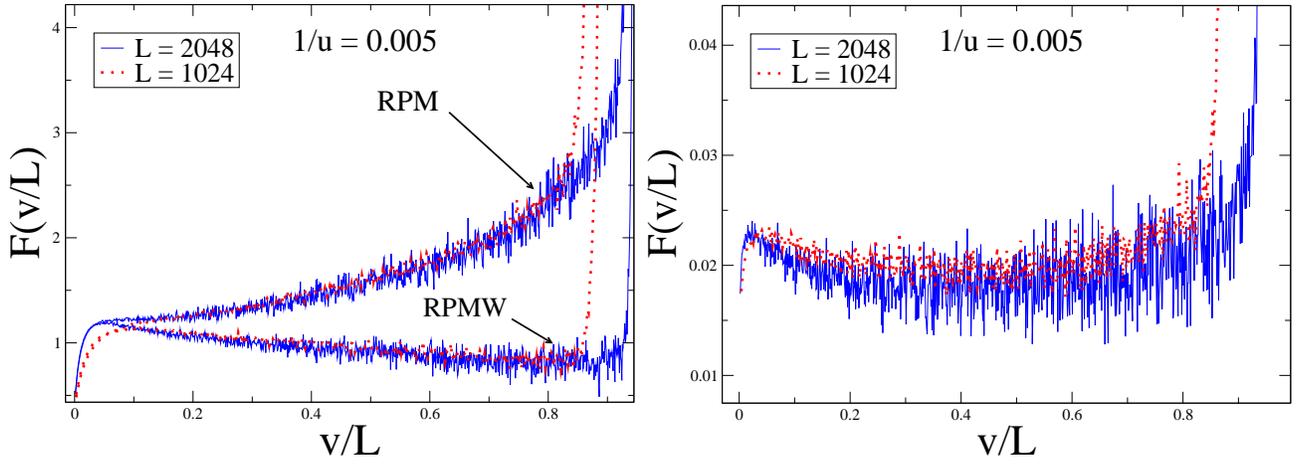

  \begin{tabular}{cc}
    \includegraphics[scale=.35]{FIGS/Fv_w005.eps} &
    \includegraphics[scale=.35]{FIGS/Fve_w005.eps} \\
  \end{tabular}
  \caption{The scaling function $F(\frac{v}{L})$ with rates $1/u=0.005$ for $v \in$ {\it odd} number with {\bf $\tau=2.0$} and {\bf $D=1.0$} (LEFT) and $v \in$ {\it even} number with {\bf $\tau=1.0$} and {\bf $D=1.0$}}
  \label{fig:FVw005}
\end{figure}

Results in this table show that the critical exponents for an {\it odd}
number of tiles removed remains unchanged by adding a wall. However, we
found that for  an {\it even} number of tiles the power-law exponent $\tau$ decreases by about one. We
also found an interesting effect on the finite-size scaling function with the addition of the wall. Fig.~\ref{fig:FVu1}
shows the scaling function for RPM and RPMW for u=1, where we  use $\tau=3.0$ and $\tau=2.0$ respectively, 
for {\it odd} number of tiles and {\it even} number of tiles, while D is kept fixed at $D=1$. 
Fig.~\ref{fig:FVw005} shows a similar plot for 1/u=0.005. The data collapse for large lattices on these plots confirms 
the FSS form (Eqn. \ref{eqn:FSS}).

\section{Conjectures for Probabilities}\label{sec:conj}
Simple conjectures for the probabilities of absorption, desorption, and reflection can be written down by considering
the rate of change between the different states.  
The probability to loose (or gain) v tiles can be written as \cite{gier}:
\be P(v,L) = \sum_{\eta \neq \eta'} \delta(v(\eta')-v(\eta)-v) w_{\eta' \rightarrow \eta} P_{\eta'}/\langle 0|0 \rangle \ee
where $P_{\eta'}/\langle 0|0 \rangle$ is {\it normalized} probability to be in the state $|v(\eta') \rangle$ (see Eqn. \ref{eqn:statst}) and $w_{\eta' \rightarrow \eta}$ is the {\it transition rate} $w_{\eta' \rightarrow \eta}$ from state $|v(\eta') \rangle$ to state $|v(\eta) \rangle$. For u=1, the normalization $\langle 0|0 \rangle$ is given by Eqn.(\ref{eqn:rpmN})
and Eqn.(\ref{eqn:rpmwN}) for RPM and RPMW, respectively.
The probability of absorption ($P_a$) for a given L has been given by Alcaraz {\it et al} for RPMW \cite{Alcaraz2}. 
In this section we will present new conjectures for the probabilities of desorption ($P_d$) and reflection ($P_r$) for RPMW.
Probabilities for RPM where first reported in \cite{gier}. These expressions results in quotients of parabolas as seen in table~\ref{table:Probs}. 
Notice the denominator for the probabilities are different with the addition of the wall since normalization 
expressions for $\langle 0|0 \rangle$ (Eqns. \ref{eqn:rpmN} and \ref{eqn:rpmwN}) are different for the cases
with and without a wall \cite{Alcaraz2}.

\begin{table}[h]
  \centering
  \caption{ Conjectures for the probabilities of an absorption event $P_a \equiv P(-1,L)$, desorption event $P_d \equiv P(v>0,L)$, reflection event $P_r \equiv P(0,L)$   
  }
  \centering
  \begin{tabular}{c | c  c  c }
    Prob.\cite{gier,Alcaraz2} & $P_a$ & $P_d$ & $P_r$ \\
    \hline\\ 
    RPM ($v \in odd$ )
    & $\frac{3L(L-2)}{4(2L+1)(L-1)}$ & $\frac{2(L-2)(L+2)}{4(2L+1)(L-1)}$ & --- \\ 
    \hline 
      RPM ($v \in even$ )
      & --- & --- & $\frac{3L^2+2L+4}{4(2L+1)(L-1)}$ \\ 
      \hline \\ 
      RPMW ($v \in odd$ )
      & $\frac{6L^2+8L-5}{4(2L+1)(2L+3)}$  & $\frac{4L^2+5L+9}{4(2L+1)(2L+3)}$  &  ---\\
      \hline 
      RPMW ($v \in even$)
      & ---  & $\frac{1.795L-4.562}{4(2L+1)(2L+3)}$  & $\frac{6L^2+17.211L+12.50}{4(2L+1)(2L+3)}$ \\
      \label{table:Probs}
  \end{tabular}
\end{table}

It is interesting to note that although this is a non-equilibrium system, simple expressions for probabilities 
can be obtained. The mean size of an avalanche $\langle v \rangle$ in the stationary state can 
be conjectured to be given by a mean-field expression \cite{alcaraz}.

\begin{equation}
  \langle v \rangle_L = P_a(L)/P_d(L)
\end{equation}

\begin{figure}[ht]
  \centering
  \vspace{0mm}\hspace{0mm}\includegraphics[scale=1.3]{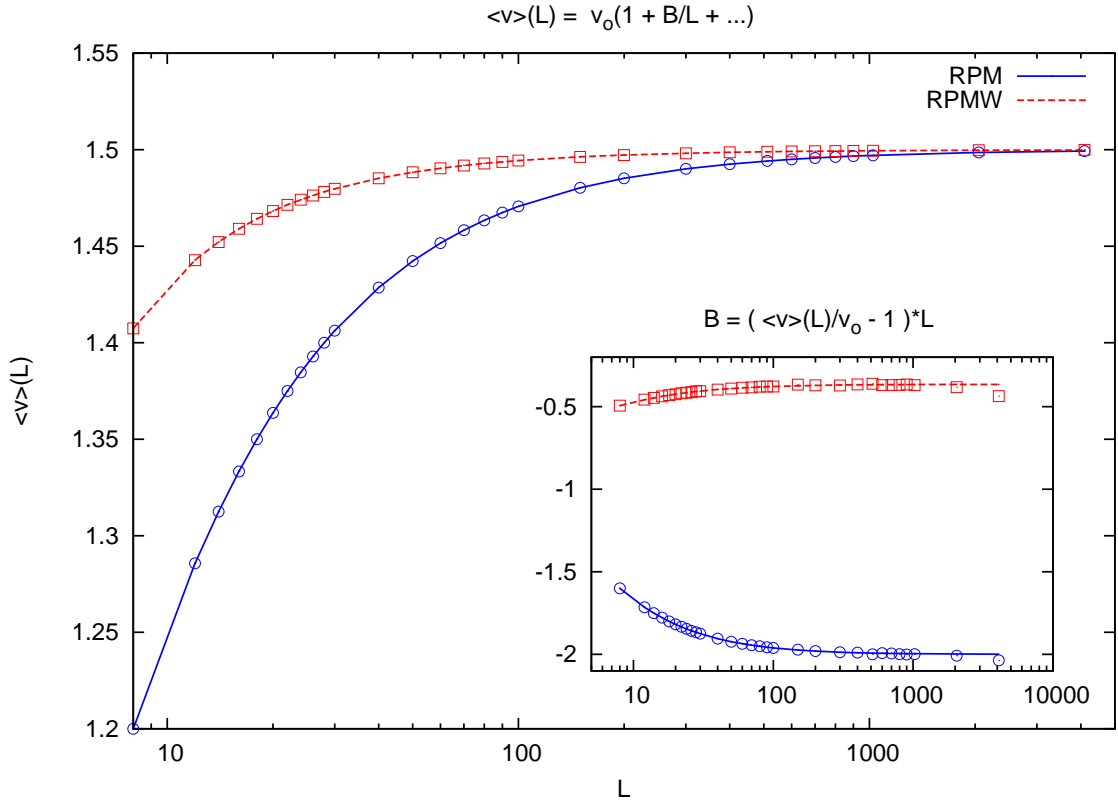}
  \caption{Average avalanche size $\langle v \rangle(L)$ plotted from the ratio of $P_a(L)/P_d(L)$ (lines) 
    compared to Monte Carlo data for {\bf RPM} ($\bigcirc$) and {\bf RPMW} ($\square$)}
  \label{fig:Vavg}
\end{figure}  

This is a quotient of parabolas, and in the large limit ($L \gg 1$)
\begin{align}
  \frac{P_a(L)}{P_d(L)} & = \frac{\alpha L^2 + \beta L + \gamma}{a L^2 + b L + c} \nonumber \\
  & = \frac{\alpha L^2 + \beta L + \gamma}{a L^2 (1  + \frac{b}{aL} + \frac{c}{aL^2})} \nonumber \\
  & \approx (\frac{\alpha}{a} + \frac{\beta}{aL} + \mathcal{O}(L^{-2}))(1  - \frac{b}{aL} + \mathcal{O}(L^{-2}))\nonumber \\  & \approx (\frac{\alpha}{a} + \frac{\beta}{aL} - \frac{\alpha b}{a^2 L} + \mathcal{O}(L^{-2})) \nonumber \\ 
  & = (\frac{\alpha}{a} + \frac{1}{aL} (\beta - \frac{\alpha b}{a}) ) \nonumber \\ 
  & = \frac{\alpha}{a} (1  + \frac{1}{L} (\frac{\beta}{\alpha} - \frac{b}{a}))  \nonumber \\ 
\end{align}

We assume that $\alpha \ne 0$ and $a \ne 0$ and dropped terms of the order of $\mathcal{O}(L^{-2}))$.
From Table \ref{table:Probs} we see that for RPM we have
\begin{align}
  \frac{P_a(L)}{P_d(L)} & = \frac{3}{2} (1  - \frac{1}{L} (\frac{6}{3} - \frac{0}{2}))  \nonumber \\ 
  \Rightarrow \langle v \rangle_L  & = \frac{3}{2} (1  - \frac{2}{L} )  \nonumber \\ 
\end{align}
while for RPMW we have:
\begin{align}
  \frac{P_a(L)}{P_d(L)} & = \frac{6}{4} (1  + \frac{1}{L} (\frac{8}{6} - \frac{6.795}{4}))  \nonumber \\ 
  \Rightarrow \langle v \rangle_L  & = \frac{3}{2} (1  - \frac{0.365}{L} )  \nonumber \\ 
\end{align}

The estimated average from the ratio of the conjectures agrees quite well as can be seen in Fig~\ref{fig:Vavg}.
This remarkable since this describes a non-equilibrium system.

It is interesting to see that the leading term on this expansion is
universal, whereas the correction term depends on the details of the model
(i.e. whether there is a wall or not). This is a similar behavior as in
finite--size scaling of the concentration of particles in certain
reaction--diffusion systems \cite{birgit1,birgit2,birgit3} and surface
exponent corrections to quantum chains using different types of boundary
conditions \cite{alcaraz_surface,Turban,BX}.

\begin{figure}[t]
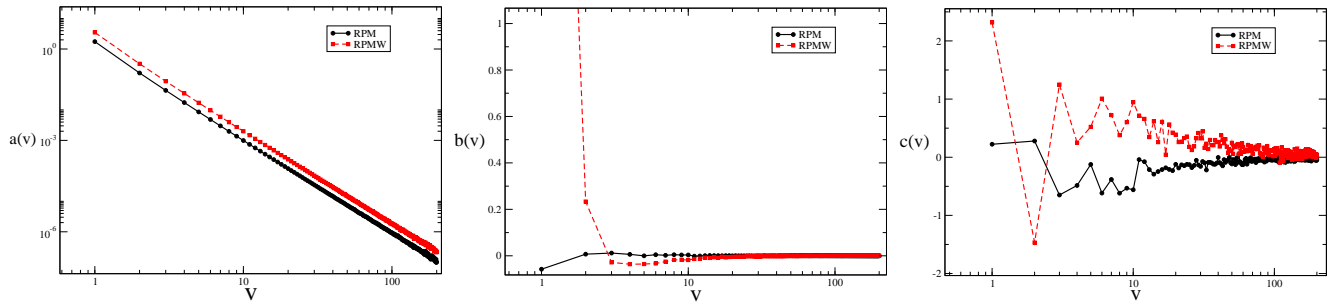

\par\vspace{3mm}
  \begin{tabular}{ccc}
    \includegraphics[scale=.235]{FIGS/av.eps} &   
    \includegraphics[scale=.235]{FIGS/bv.eps} &
    \includegraphics[scale=.235]{FIGS/cv.eps} 
    \\
  \end{tabular}
  \caption{Behavior of the quadratic term $a(v)$, linear term $b(v)$ and constant term $c(v)$ obtained from fits to 
Eqn.~\ref{eqn:Pv1} and Eqn.~\ref{eqn:Pv2}.}
  \label{fig:Vs}
\end{figure}

The simple functional form for the probabilities shown in table \ref{table:Probs} suggests we can guess a
general quadratic expression in L for the probabilities P(v,L) by fixing the denominator as:
\begin{equation}\label{eqn:Pv1}
  P_{RPM}(v,L)=\frac{a(v) L^2 + b(v) L + c(v)}{4(2L+1)(L-1)}
\end{equation}
\begin{equation}\label{eqn:Pv2}
  P_{RPMW}(v,L)=\frac{a(v) L^2 + b(v) L + c(v)}{4(2L+3)(2L+1)} 
\end{equation}
and fitting for the parameters \{$a(v)$, $b(v)$, $c(v)$\} in the  forms
(\ref{eqn:Pv1}) and (\ref{eqn:Pv2}). The behavior of these parameters with
respect to $v$ is shown in Fig~\ref{fig:Vs}. As expected, the quadratic term shows a power-law behavior $\propto v^{-3.0}$. The linear and constant terms however do show different behavior, but we were not able to reduce it to an analytical form. 
Consistency on the fits demands that 
(see table~\ref{table:Probs})
\begin{equation}\label{eqn:Psv1}
  \sum_{v>0} P_{RPM}(v,L)=\frac{2 L^2 + 0 L -8 }{4(2L+1)(L-1)} 
\end{equation}
\begin{equation}\label{eqn:Psv2}
  \sum_{v>0}P_{RPMW}(v,L)=\frac{4 L^2 + 5 L + 9}{4(2L+3)(2L+1)} 
\end{equation}

The sums over the parameters $\sum_v$ \{$a(v), b(v), c(v)$\} are shown in Table \ref{table:fitsParabolas} for RPM and RPMW.
We see that the quadratic $a(v)$ and linear $b(v)$ terms capture the
behavior in $v$ quite well, whereas as shown in Fig \ref{fig:Vs} the constant term $c(v)$ is dominated by fluctuations on the fits.
\begin{table}[ht]
  \centering
  \begin{tabular}{c | c  c  c }
    & $\sum_v a(v)$ & $\sum_v b(v)$& $\sum_v c(v)$ \\
    \hline\\ 
    RPM     & 1.99999 & -0.00375 & -13.8002 \\ 
    RPMW    & 3.99997 &  4.96604  & 30.4445 \\ 
  \end{tabular}
  \caption{ Sums over $v$ for the parameters resulting from the fits}
  \label{table:fitsParabolas}
\end{table}

\section{conclusion}

We studied the non-equilibrium statistical model known as the raise and peel model.
We have confirmed that this model retains several features as predicted from conformal invariance
for  stochastic profiles characterizing the system when changing the boundary conditions.
We allowed one boundary to fluctuate and demonstrated 
that the temporal profile of a stochastic quantity follows its expected behavior from stochastic dynamics
where the long time behavior is dominated by the lowest non-zero eigenvalue. 
We study the surface dynamics by looking at avalanche distributions exhibiting power-law distributions.
Using the finite-size scaling formalism we confirm the universality exponent $\tau=3.0$ for the Raise and Peel
with different boundary conditions and we identified an even/odd effect with a new exponent $\tau=2.0$ for avalanches 
with an even number of tiles removed. We also found new conjectures for the probability of desorption and reflection with
a wall added to the system and checked that they agree with Monte Carlo data.

\section*{Acknowledgements}
We would like to thank M. Henkel and the Groupe de Physique Statistique at Nancy University for helpful discussions. B. W.-K. thankfully acknowledges support from the NSF  under the grant PHY-0969689.

\bibliography{mybib_nourl}{}
\bibliographystyle{unsrt}

\end{document}